\documentclass[aps,prd,twocolumn,groupedaddress,showpacs,nofootinbib,amssymb,longbibliography]{revtex4-2}
\usepackage[dvips]{graphicx}
\usepackage{amssymb}
\usepackage{amsmath}
\usepackage{gensymb}

\usepackage{graphicx,color}
\usepackage{amsfonts}
\usepackage{bm}
\usepackage{cancel}
\usepackage{comment}
\usepackage{braket}
\usepackage{physics}
\usepackage{float}

\usepackage[colorlinks=true, allcolors=blue]{hyperref} 
\usepackage{amsmath}
\usepackage{ulem} 
\usepackage{caption}
\usepackage{subcaption}
\usepackage{url}
\usepackage{natbib}
\usepackage{soul}

\usepackage[colorlinks=true, allcolors=blue]{hyperref}

\usepackage{graphicx}

\usepackage{subcaption}
\usepackage{hyperref}

\newcommand\be{\begin{equation}}
\newcommand\ee{\end{equation}}

\DeclareUnicodeCharacter{2212}{-}
\allowdisplaybreaks[4]

\begin{document}

\bibliographystyle{apsrev4-2}

\tolerance=5000
\title{Impact of correlated magnetic noise on directional stochastic \\
gravitational-wave background searches}
\author{Stavros Venikoudis$^{1}$}
\email{stavros.venikoudis@uclouvain.be}
\author{Federico De Lillo$^{1, 2}$}
\email{federico.delillo@uclouvain.be}
\author{Kamiel Janssens$^{2,3,4,5}$}
\email{kamiel.janssens@uantwerpen.be}
\author{Jishnu Suresh$^{1,3}$}
\email{jishnu.suresh@uclouvain.be}
\author{Giacomo Bruno$^{1}$}
\email{giacomo.bruno@uclouvain.be}

\affiliation{$^{1}$Centre for Cosmology, Particle Physics and Phenomenology (CP3), Universite catholique de Louvain, Louvain-la-Neuve, B-1348, Belgium}
\affiliation{$^{2}$Universiteit Antwerpen, Prinsstraat 13, 2000 Antwerpen, Belgium}
\affiliation{$^{3}$Universit\'e C$\hat{o}$te d’Azur, Observatoire C$\hat{o}$te d’Azur, CNRS, Artemis, 06304 Nice, France}
\affiliation{$^{4}$Department of Physics, The University of Adelaide, Adelaide, SA 5005, Australia}
\affiliation{$^{5}$ARC Centre of Excellence for Dark Matter Particle Physics, Australia}

\begin{abstract}

One potential factor that could impede searches for the stochastic gravitational-wave background (SGWB), arising from the incoherent superposition of a multitude of weak and unresolvable gravitational-wave signals in the Universe, is correlated magnetic noise at Earth-scale distances, such as the Schumann resonances. As the sensitivity of terrestrial detectors to SGWBs increases, these effects are expected to become even more pronounced, emphasizing the importance of considering correlated noise sources in SGWB searches. In this study, we explore for the first time the impact of this class of noise on anisotropic SGWB searches. We focus on the low-frequency range of $20-200$ Hz with a resolution of 1/32 Hz, using data from the LEMI-120 and Metronix MFS-06e magnetometers, located at the sites of the Advanced LIGO and Advanced Virgo interferometers, respectively, during the third observing run. These measurements were then compared with the directional upper limit derived from gravitational-wave data collected during the third observing run of the LIGO-Virgo-KAGRA collaboration. The results indicate that the magnetic noise correlations are below the interferometers' sensitivity level during the O3 observing run. The potential impact of magnetic correlations on the Advanced LIGO detectors has also been investigated at the A+ design sensitivity level.
\end{abstract}

\maketitle

\section{Introduction}
As of today, the LIGO-Virgo-KAGRA (LVK) collaboration \cite{KAGRA:2013rdx,aLIGO:2020wna,Tse:2019wcy,LIGOScientific:2014pky,Virgo:2019juy,VIRGO:2014yos,Aso:2013eba,KAGRA:2020tym} has cataloged 90 gravitational-wave (GW) events \cite{LIGOScientific:2018mvr,LIGOScientific:2020ibl,KAGRA:2021vkt} from the first three observing runs O1, O2, and O3. These signals are consistent with compact binary coalescences (CBCs). With the ongoing fourth observing run (O4), one detection of CBC event is being reported every few days. This observing run may lead to the first detection of other kinds of GW signals, such as quasi-monochromatic GWs, short-duration unmodeled GW transients, as well as persistent and unmodeled GWs, namely stochastic gravitational wave backgrounds (SGWBs) \cite{Romano:2016dpx,Renzini:2022alw, Christensen:2018iqi,vanRemortel:2022fkb}, where the adjective stochastic refers to the GW strain of the signal being a random variable. 
A SGWB results from the superposition of individually undetectable GW signals, and its detection would be an observational milestone of GW astronomy with ground-based GW interferometers.

An SGWB signal can originate from either the primordial Universe or astrophysical sources. An SGWB from cosmological phenomena \cite{Caprini:2018mtu}, such as the inflationary era \cite{Turner:1996ck,Easther:2006gt} of the primordial Universe, first-order phase transitions \cite{Marzola:2017jzl}, cosmic strings \cite{Kibble:1976sj,Sarangi:2002yt,Damour:2004kw,Siemens:2006yp}, domain walls \cite{Martin:1996ea}, and primordial black holes \cite{Mandic:2016lcn} could provide unique insights into the evolution of the Universe, that are inaccessible by any other means. This kind of SGWB is expected to be predominantly isotropic, similar to the Cosmic Microwave Background \cite{1965ApJ...142..419P}. An astrophysical SGWB from isolated neutron stars \cite{Rosado:2012bk, ellipticity_Talukder:2014eba, ellipticity_DeLillo:2022blw, pulsar_glitch_DeLillo:2022dau}, compact binary coalescences \cite{LIGOScientific:2021qlt,Zhu:2011bd}, core-collapse supernovae \cite{Buonanno:2004tp}, and stellar core collapses \cite{Crocker:2015taa} could provide new insights in astrophysical populations and evolution processes. In contrast to a cosmological SGWB, an astrophysical SGWB in the nearby Universe follows the distribution of the matter in space, such as galaxies or galaxy clusters, and is likely to show a degree of spatial anisotropy.

To target SGWB anisotropies from localized sources with a network of ground-based GW detectors, one cross-correlates the data streams from different detector pairs \cite{Allen:1997ad,Romano:2016dpx}.
The principle of radiometry, which takes advantage of the Earth's rotation \cite{Ballmer:2005uw,Mitra:2007mc}, is then employed to produce sky-maps of the GW power in the sky \cite{KAGRA:2021mth,KAGRA:2021rmt}. 
Radiometer searches for SGWB anisotropies can either target narrow-band sources in a specific sky direction or broad-band sources emitting GW over a wide frequency band and are known in the literature as narrow-band radiometer and broad-band radiometer (BBR) searches \cite{KAGRA:2021mth}.
Recently, an all-sky, all-frequency (ASAF) radiometer analysis was made available to produce sky maps of GW power at any frequency \cite{KAGRA:2021rmt}. However, none of these searches
has detected an SGWB yet, instead setting upper limits on its strength \cite{KAGRA:2021mth,KAGRA:2021rmt}.


These analyses assume that there is no correlated noise source between the detectors. However, multiple studies have shown that persistent global magnetic interferences \cite{Thrane:2013npa,Janssens:2022tdj} can couple to the GW interferometers, giving rise to a correlated noise among the detectors of the network. Even though no impact is currently observed, projections predict significant coupling for Virgo and LIGO at A+ design sensitivity \cite{A+,Janssens:2022tdj}, as well as for future detectors such as the Einstein Telescope \cite{Janssens:2021cta}. Such correlated noise in GW detectors is a non-negligible obstacle for detecting and characterizing the SGWB.

A significant source of correlated magnetic noise is the Schumann resonances \cite{Schumann+1952+149+154, Schumann+1952+250+252, Jackson:1998nia}. These resonances are electromagnetic excitations inside the conducting cavity between the Earth and the ionosphere, acting as a spherical waveguide. Atmospheric discharges, such as cloud-to-ground lightning strikes worldwide, generate them. 
The spectrum of the Schumann resonances presents a series of broad spectral peaks ranging in the 8-60 Hz frequency band, whose amplitude varies depending on the season and location \cite{Meyers:2020qrb}. 
In addition to Schumann resonances, correlations due to magnetic fields can also occur at higher frequencies, originating from the superposition of individual lightning strikes propagating over long distances on Earth before degrading \cite{Janssens:2022tdj}.

In the near future, as the GW detectors approach their design sensitivities, correlated magnetic noise may no longer be negligible when performing SGWB searches.
Many Wiener filtering-based studies have been developed to subtract this noise \cite{Thrane:2014yza,Coughlin:2016vor,Coughlin:2018tjc}. In addition to that, the analytical modeling of the correlated noise to assess the impact on the detectors has been extensively studied \cite{Himemoto:2017gnw, Himemoto:2019iwd}. \citet{Meyers:2020qrb} also proposed to jointly estimate SGWB and correlated magnetic noise within a Bayesian framework. Alternatively, Fisher-matrix approaches are available in the literature, such as in \citet{Himemoto:2023keu}.

In this paper, we focus for the first time on the impact of correlated magnetic noise on searches for anisotropic SGWB with Earth-based detectors. To assess the potential contamination of these searches by correlated magnetic noise, we separately measure this noise. Then, we compare this term with the ASAF strain upper limits, as well as the BBR upper limits of GW flux obtained from the O3 run of the LIGO-Virgo-KAGRA collaboration \cite{KAGRA:2021rmt,KAGRA:2021mth}.
We also perform a comparison between these measurements and the sensitivity of the ASAF directional search, once the LIGO detectors reach the A+ design configuration \cite{A+,Cooper:2022jfr}.

The structure of the paper is as follows. In Section \ref{sec:CC_with_correlated_noise}, we outline the cross-correlation statistic in the presence of correlated magnetic noise. In Section \ref{sec:magnetic_noise_measurement}, we discuss the instruments used to measure the noise, the pre-processing of data applied in this analysis and the coupling mechanism of magnetic correlations with the GW interferometers. In Section \ref{sec:directional_searches_with_correlated_magnetic_noise}, we develop the formalism to evaluate the impact of correlated magnetic noise on directional SGWB searches
. In Section \ref{sec:results}, we present the main results of this work, namely the estimates of the effective SGWB spectra that arise from the magnetic noise in the current ground-based GW detector network for different SGWB directional searches, which we call \textit{magnetic budgets}.
Concluding remarks are also provided. In Appendix \ref{sec:appendix_ASAF_HL_HV_LV}, we present the effective GW strain for each sky direction, related to correlated magnetic terms for each individual baseline of the LVK detector network. Lastly, in Appendix \ref{sec:appendix_A+_projection}, we perform a projection of an \textit{power-law integrated-like (PI-like) sensitivity curve at each sky direction} for the ASAF directional search, to explore the impact of magnetic correlations on LIGO detectors once they reach the A+ design sensitivity levels, followed by the derivation of the method in appendix \ref{sec:appendix_PI_curve}.

\section{Cross-Correlation statistic IN THE PRESENCE OF CORRELATED MAGNETIC NOISE}
\label{sec:CC_with_correlated_noise}

Anisotropic SGWB searches aim at characterizing the angular distribution of the dimensionless GW energy density parameter $\Omega_{\mathrm{gw}}(f,\boldsymbol{\hat{n}})$ \cite{KAGRA:2021rmt,LIGOScientific:2019gaw,KAGRA:2021mth,Romano:2016dpx,Thrane:2009fp},

\begin{equation}
\centering
\label{Omega_mag}
\Omega_{\mathrm{gw}}(f,\boldsymbol{\hat{n}})
=\frac{1}{\rho_c}\frac{\mathrm{d}^3\rho_{\rm gw}}{\mathrm{d}({\mathrm{ln}f}) \mathrm{d}^{2}{\boldsymbol{\hat{n}}}}=\frac{2 \pi^2}{3H_0^2}f^3 \mathcal{P}_{\rm gw}(f,\boldsymbol{\hat{n}}),
\end{equation}
where $d\rho_{\rm gw}$ denotes the GW energy density per logarithmic frequency unit $\mathrm{d}(\mathrm{ln}f)$. The quantity $\rho_c$ expresses the critical energy density for a spatially flat Universe, with $H_0 = 67.9 \rm km s^{-1} Mpc^{-1}$ being the Hubble parameter today according to the Planck 2015 measurements \cite{Planck:2015fie} and $\mathcal{P}_{\rm gw}(f,\boldsymbol{\hat{n}})$ characterizes the angular power spectrum of the SGWB. Under the assumption of a Gaussian, unpolarized and stationary GWB, the one-sided $\mathcal{P}_{\rm gw}(f,\boldsymbol{\hat{n}})$ is related to the expectation value of the GW strain distribution $h_A(f,\boldsymbol{\hat{n}})$ across different sky directions as \cite{Romano:2016dpx}
\begin{equation}
\centering
\langle h_A^*(f, \boldsymbol{\hat{n}}) h_{A'}(f', \boldsymbol{\hat{n}}') \rangle = \frac{1}{4} \mathcal{P}_{\rm gw}(f, \boldsymbol{\hat{n}}) \delta_{AA'} \delta(f - f') \delta(\boldsymbol{\hat{n}}, \boldsymbol{\hat{n}}'),
\end{equation}
where the index A expresses the GW polarization index ($\{\rm +, \rm x\}$), and the asterisk denotes the complex conjugate of the GW strain amplitude.

The SGWB signal is expected to be both weak and inherently random. As a result, distinguishing this type of signal from instrumental noise on a single detector poses a significant challenge. To address this, the statistical method commonly used in SGWB searches is the cross-correlation \cite{Allen:1997ad,Romano:2016dpx} of data streams from a detector pair. However, the detectors' output can also contain contributions from correlated noise. In this work, we will consider correlated noise from magnetic origin, which exhibits correlations over Earth-scale distances \cite{Janssens:2022tdj}.

Considering two GW detectors $I$ and $J$, the output of each of them is given as a time-series $s_{I}(t)$ and $s_{J}(t)$. This output is more convenient to be expressed in the frequency domain as, 
\begin{align}
\centering
\label{output}
\tilde{s}_{I}(f) &= \tilde{h}_{I}(f) + \tilde{n}^{\rm indep}_{I}(f) + \tilde{n}^{\rm corr}_{I}(f), \\
\tilde{s}_{J}(f) &= \tilde{h}_{J}(f) + \tilde{n}^{\rm indep}_{J}(f) + \tilde{n}^{\rm corr}_{J}(f),
\end{align}
where $\tilde{h}_{I}(f)$, $\tilde{h}_{J}(f)$ indicate the strain amplitudes of the signal observed in detectors $I$ and $J$, $\tilde{n}^{\rm indep}_{I}(f)$ and $\tilde{n}^{\rm indep}_{J}(f)$ denote any kind of independent noise namely, the intrinsic noise of the detectors, as well as other sources of uncorrelated environmental noise. Lastly 
$\tilde{n}^{\rm corr}_{I}(f)$, $\tilde{n}^{\rm corr}_{J}(f)$ represent the correlated noise terms \cite{AdvLIGO:2021oxw,Virgo:2022ypn,Fiori:2020arj}. In this model, we 
focus only on magnetic correlations $\tilde{n}^{ \mathrm{mag}}_{I}(f), \, \tilde{n}^{ \mathrm{mag}}_{J}(f)$ and ignore other sources of correlated noise. 
After cross-correlating $\tilde{s}_{I}(f)$ and $\tilde{s}_{J}(f)$, the expectation value is
\begin{equation}
\label{expected_value}
\centering
\begin{split}
\langle \tilde{s}^{*}_{I}(f) \tilde{s}^{}_{J}(f') \rangle = \langle \tilde{h}^{*}_{I}(f) \tilde{h}^{}_{J}(f') \rangle + \cancel{\langle \tilde{n}^{* \mathrm{indep}}_{I}(f) \tilde{n}^{{\mathrm{indep}}}_{J}(f') \rangle} \\
+ 
\langle \tilde{n}^{* \mathrm{mag}}_{I}(f) \tilde{n}^{{\mathrm{mag}}}_{J}(f') \rangle ,
\end{split}
\end{equation}
where the term $\langle \tilde{n}^{* \mathrm{indep}}_{I}(f) \tilde{n}^{{\mathrm{indep}}}_{J}(f') \rangle$ vanishes, because there is no correlation of instrumental noise, 
due to the geographical separation of detectors. 

One can investigate the potential impacts of correlated magnetic noise on SGWB searches by measuring the term $\langle \tilde{n}^{* \mathrm{mag}}_{I}(f) \tilde{n}^{{\mathrm{mag}}}_{J}(f') \rangle $ in Eq. \eqref{expected_value}
, which may bias the cross-correlation statistic if it is comparable with $\langle \tilde{h}^{*}_{I}(f) \tilde{h}^{}_{J}(f') \rangle$.

\section{Measuring the magnetic noise}
\label{sec:magnetic_noise_measurement}

We investigate the presence of correlated magnetic noise using the magnetometers located outside the detector sites, the two LEMI \cite{lemi} magnetometers at each LIGO site and the two Metronix MFS-06e magnetometers \cite{metronix} at Virgo. Each of the two magnetometers at every site is labelled with the index $i$ or $j$. The LEMI magnetometers are aligned with the X- and Y-arms of the LIGO detectors, while the Metronix magnetometers are oriented along the North-South and East-West directions.


We analyze the O3 data of these magnetometers for the Hanford (H), Livingston (L), and Virgo (V) detector networks. 
We obtain the cross-correlation spectra of each magnetometer pair $I_iJ_j$ for every GW detector baseline $IJ$ ($I$, $J$ =\{H, L, V\}) as an output of the \texttt{Stochmon} (Stochastic Monitor) data-quality tool \cite{stochmon}. 
Following \cite{KAGRA:2021kbb}, we downsample the magnetometer time-series $m_{I_{i}}$ to 512 Hz and apply a high-pass filter to them using a Butterworth filter with a knee frequency of 11 Hz. Then, we divide each data stream to segments $m_{I_{i}t}$ of duration $\tau=1000$ sec, labeling each segment with $t$. The data streams are Hann-windowed and 50\%-overlapped. Afterward, we apply discrete short Fourier transform $\tilde{m}_{I_{i}t}(f)$ on each of these segments and eventually coarse-graining the resulting spectra to a $1/32$ Hz frequency resolution.

Finally, we evaluate the cross-correlation spectra $(\tilde{m}^{*}_{I_{i}t}\tilde{m}_{J_{j}t})(f)$, average them over the segments and define the cross-spectral density for the correlated magnetic noise $\Tilde{m}^{\mathrm{mag}}_{IJ}(f)$ of each detector pair as
\begin{align}
\centering
\label{magnetic_CSD}
\Tilde{m}^{\mathrm{mag}}_{IJ}(f)=&\frac{2}{\tau} \Bigg[ \abs{(\Tilde{m}^{*}_{{I}_{i}}\Tilde{m}_{{J}_{i}})(f)} ^2+\abs{(\Tilde{m}^{*}_{{I}_{i}} \Tilde{m}_{{J}_{j}})(f)} ^2 \nonumber \\
&+\abs{(\Tilde{m}^{*}_{{I}_{j}}\Tilde{m}_{{J}_{i}})(f)}^2+ \abs{(\Tilde{m}^{*}_{{I}_{j}}\Tilde{m}_{{J}_{j}})(f)}^2\Bigg]^{1/2},
\end{align}
where we stay conservative by considering the quadratic sum of the four pairs of magnetometers and discard the phase information, as in \cite{Janssens:2022tdj}. 

Magnetic noise can couple to the GW interferometers, resulting in an additional source of noise for the GW strain in the data. To correctly measure the extra noise strain spectral density from correlated magnetic noise in SGWB searches, it is important to understand how this coupling mechanism works at different GW detector sites.

The coupling mechanism varies for different frequency ranges. Specifically, in the 20-100 Hz band, global magnetic fields, mainly Schumann resonances, induce forces on the magnets located on the electromagnetic actuators of the test masses and on suspension systems as well \cite{Thrane:2013npa,LSC:2018vzm}. Other coupling mechanisms may be possible at higher frequencies, such as the interaction with the cabling system \cite{Janssens:2022tdj}. To quantify the coupling of the magnetic field to the GW-strain channels, one has to measure the \textit{magnetic coupling function} $\kappa_{I}(f)$ for the detector $I$ \cite{Janssens:2022tdj} (the same applies for the detector $J$), which is considered as the product of two time-independent quantities\footnote{In reality the coupling function changes to some extent over long observation periods. However, as shown in \cite{Janssens:2022tdj}, any uncertainty related to time fluctuations is second order, and hence, we can ignore any time dependence in this work.}

\begin{equation}
\centering
\label{coupling_total}
\kappa_{I}(f)= \kappa_{{I}, \mathrm{OTI}} (f)  \cdot   \kappa_{{I}, \mathrm{ITC}}(f),
\end{equation}
where $\kappa_{{I}, \mathrm{ITC}}(f)$ is called inside-to-GW-channel (ITC) coupling function and $\kappa_{{I}, \mathrm{OTI}}(f)$ outside-to-inside (OTI) coupling function\footnote{The quantities we refer to as coupling functions follow the definition in Eq. (1) of \citet{AdvLIGO:2021oxw}. These quantities are generally complex and used as an alternative to transfer function under the assumption of lack of perfect coherence \cite{AdvLIGO:2021oxw}}. The key coupling features are captured in the ITC coupling function, while the OTI coupling provides a level of shielding/amplification caused by the building.

It is possible to estimate $\kappa_{{I}, \mathrm{ITC}}(f)$ by using the sensors that are inside the detector building and monitoring the fluctuations in the magnetic field at the detector site \cite{AdvLIGO:2021oxw,Fiori:2020arj}. 
Given the weakness of $\kappa_{{I}, \mathrm{ITC}}(f)$ and the environmental magnetic field (of the order of pT), one performs intense magnetic injections (of the order of nT) using induction coils in different buildings to achieve measurable effects on the GW strain channels.
Conducting such injections at different locations inside the detectors' buildings and selecting the maximum value of the coupling at each frequency resulted in a sitewide spectrum of $\kappa_{{I}, \mathrm{ITC}}(f)$ for the two LIGO interferometers \cite{AdvLIGO:2021oxw}.
Similar magnetic injections happened weekly for the Virgo detector, mainly in its central building, throughout O3, resulting either in measurements or upper limits on $\kappa_{{I}, \mathrm{ITC}}(f)$ \cite{Fiori:2020arj}.

The OTI function $\kappa_{{I}, \mathrm{OTI}}(f)$ in Eq. \eqref{coupling_total} has been considered equal to unity in \cite{LIGOScientific:2019vic,KAGRA:2021kbb}, following the measurements of coil-generated fields at the Hanford site. More recently, the reports about lighting activity near the Livingston interferometer have made it possible to obtain $\kappa_{{I}, \mathrm{OTI}}(f)=0.7^{+0.4}_{-0.3}$ \footnote{Based on the observations of the OTI magnetic coupling function in \cite{Janssens:2022tdj} for the Livingston detector, values larger than one were observed in a small subset of cases. For a more detailed discussion, we refer the interested reader to Appendix A of \cite{Janssens:2022tdj}.}
for the magnetic field of the order of nT \cite{Janssens:2022tdj}. 
We adopt the latter value in this work and assume the same value for LIGO Hanford and Virgo.

\section{Directional searches and correlated magnetic noise}
\label{sec:directional_searches_with_correlated_magnetic_noise}
When searching for the anisotropies of SGWB in particular directions in the sky with ground-based GW detectors \cite{Ballmer:2005uw,Mitra:2007mc,Thrane:2009fp}, one employs cross-correlation techniques with the inclusion of a time-varying phase delay. This delay expresses the required time difference of a GW signal to arrive at the detectors coming from a specific sky direction. Thus, by cross-correlating the data, GW signals from specific directions in the sky will interfere constructively. Following this formalism, the radiometer map-making at every frequency bin can be performed. Assuming to be dealing with point-like sources, one expands the SGWB angular power spectrum in the \textit{pixel basis}

\begin{equation}
\centering
\mathcal{P}_{\rm gw}(f,\boldsymbol{\hat{n}})=\sum_p \mathcal{P_{\rm gw}}_p(f) e_p(\boldsymbol{\hat{n}}),
\end{equation}
where $e_p(\boldsymbol{\hat{n}})=\delta^2(\boldsymbol{\hat{n}}-\boldsymbol{\hat{n}}_p)$ and $\boldsymbol{\hat{n}}_p$ indicates the direction of a pixel p. Following the prescription of the radiometer analysis,
one can derive the maximum-likelihood estimator for $\mathcal{P}_{\rm gw}(f,\boldsymbol{\hat{n}})$ as
\begin{equation}
\centering
\label{estimator}
\boldsymbol{\hat{\mathcal{P}}}_{\rm gw}(f) = \boldsymbol{\Gamma}^{-1}_{\rm gw}(f) \vdot \boldsymbol{X}_{\rm gw}(f),
\end{equation}
where $\boldsymbol{X}_{\rm gw}$ and $\boldsymbol{\Gamma}_{\rm gw}$ are called \textit{dirty map} and \textit{Fisher matrix} and are defined as
\begin{equation}
\centering
\label{Dirty}
X^{IJ}_{{\rm gw}_p}(f)= 2 \Delta f \mathfrak{Re} \sum_{t} 
\frac{\gamma^{IJ*}_{ft,p}\tilde{s}^{*}_{I}(t;f)\tilde{s}_{J}(t;f)}{P_{{I}}(t;f)P_{{J}}(t;f)},
\end{equation}
\begin{equation}
\centering
\label{Fisher}
\Gamma^{IJ}_{{\rm gw}_{pp'}}(f)
= \frac{\tau \Delta f}{2} \mathfrak{Re} \sum_{t} 
\frac{\gamma^{IJ*}_{ft,p} \gamma^{IJ}_{ft,p'}}{P_{{I}}(t;f)P_{{J}}(t;f)},
\end{equation}
with $\tau$ the segment duration, $\Delta f$ the frequency resolution, and $P_I(t;f)$, $P_J(t;f)$ the power spectral densities of the detectors $I$ and $J$. The geometrical factor $\gamma^{IJ}_{ft,p}$, known as overlap reduction function \cite{Thrane:2009fp,Christensen:1992wi,Flanagan:1993ix,Mitra:2007mc,Finn:2008vh,Romano:2016dpx}, quantifies the combined response of a detector baseline to GW signals incorporating a time varying phase delay, and it is defined as
\begin{equation}
\centering
\label{orf}
\gamma^{IJ}_{ft,p} = \sum_A F_{I}^A(\boldsymbol{\hat{n}},t)F_{J}^A(\boldsymbol{\hat{n}},t)e^{2 \pi i f \boldsymbol{\hat{n}}\vdot \boldsymbol{\Delta x_{IJ}}(t)/c}.
\end{equation}
A is the polarization index, assuming statistical equivalence of the polarization states ($\{\rm +, \rm x\}$) \cite{Allen:1997ad,Romano:2016dpx}. $\boldsymbol{\Delta x_{IJ}}(t)$ is the separation vector between the detectors' vertices, and $F^{A}_{I/J}$ represent the antenna pattern functions in the small antenna limit \cite{Romano:2016dpx} for each detector. The uncertainty of $\mathcal{\hat{P}}_{\rm gw}(f,\boldsymbol{\hat{n}})$ is given by
\begin{equation}
\label{sigma}
\centering
\sigma_{{\boldsymbol{\hat{n}}}}(f)=
[\Gamma]^{-1/2}_{\boldsymbol{\hat{n}\hat{n}}}(f).
\end{equation}
Both Eqs. (\ref{estimator}) and (\ref{sigma}) involve the inversion of the Fisher matrix, which can be singular and therefore requires regularization to approximately solve Eq. (\ref{estimator}). However, under the assumption of point-like sources and given the current sensitivity of the detectors, one can ignore correlations between neighboring pixels \cite{Agarwal:2021gvz}. Hence, we consider the Fisher matrix to be diagonal without any significant effect on this analysis, making its inversion trivial.

Following the method described above, to quantify the impact of magnetic correlations on directional SGWB searches, we define an effective angular power spectrum $\mathcal{P}_{\rm mag}(f,\boldsymbol{\hat{n}})$ of magnetic origin that we estimate as
\begin{equation}
\centering
\label{estimator_mag}
\boldsymbol{\hat{\mathcal{P}}}_{\rm mag}(f) = \boldsymbol{\Gamma}_{\rm gw}^{-1}(f) \vdot \boldsymbol{X}_{\rm mag}(f),
\end{equation}
where we have introduced the effective
GW dirty map $X_{\rm mag}(f)$
\begin{equation}
\centering
\label{Dirty_mag}
X^{IJ}_{{\rm mag}_p}(f)= 2 \Delta f \mathfrak{Re} \sum_{t} 
\frac{\gamma^{IJ*}_{ft,p}  \abs{\kappa_{I}(f)}  \abs{\kappa_{J}(f)} 
\Tilde{m}^{\mathrm{mag}}_{IJ}(f)}{P_{{I}}(t;f)P_{{J}}(t;f)},
\end{equation}
with the coupling functions\footnote{{We stress that we are using the absolute value of the coupling function, following \cite{Janssens:2022tdj}, and hence discarding the phase information. This, together with the definition of $\Tilde{m}^{\mathrm{mag}}_{IJ}(f)$, tacitly implies that we are neglecting possible interactions of the phases of the coupling functions and $\Tilde{m}^{\mathrm{mag}}_{IJ}(f)$ with the phase of the overlap reduction function.}} ensuring that $X^{IJ}_{{\rm mag}_p}(f)$ has the same units as the regular $X^{IJ}_{{\rm gw}_p}(f)$. 
We combine the data from the three baselines, HL, HV, and LV, to estimate the effective GW angular power spectrum associated with the HLV network during O3. 
Adding more baselines significantly enhances the search sensitivity as more blind spots due to the antenna pattern functions of detectors are covered \cite{Talukder:2010yd}.  

To estimate the impact of correlated magnetic noise on searches for anisotropic SGWB, we sum the effective GW dirty maps and the Fisher matrices over all the baselines $IJ$, before evaluating the effective estimator, namely,
\begin{equation}
\centering
\label{DIRTY_TOTAL}
{X}^{\rm HLV}_{{\rm mag}}(f,\boldsymbol{\hat{n}})= 
\sum_{I} \sum_{J>I} {X}^{ IJ}_{{\rm mag}}(f,\boldsymbol{\hat{n}}),
\end{equation}

\begin{equation}
\centering
\label{FISHER_final}
{\Gamma
}^{\rm HLV}(f,\boldsymbol{\hat{n}})= \sum_{I} \sum_{J>I} {\Gamma}^{ IJ}_{{\rm gw}}(f,\boldsymbol{\hat{n}}),
\end{equation}

yielding

\begin{equation}
\centering
\label{est_final}
{\hat{\mathcal{P}}}^{\rm HLV}_{\rm mag}(f,\boldsymbol{\hat{n}}) = {\Gamma}^{\rm HLV}(f,\boldsymbol{\hat{n}})^{-1}  {X}^{\rm HLV}_{\rm mag}(f,\boldsymbol{\hat{n}}),
\end{equation}
with uncertainty,
\begin{equation}
\label{sigma_estimator_final}
\centering
\sigma^{\rm HLV}_{{\boldsymbol{\hat{n}}}}(f)=
[\Gamma^{\rm HLV}]^{-1/2}_{\boldsymbol{\hat{n}\hat{n}}}(f).
\end{equation}

%
%

\section{Results}

\begin{figure*}[] 
    \centering
    \begin{minipage}[]{0.48\textwidth}
        \centering
\includegraphics[width=\textwidth,height=3.5in]{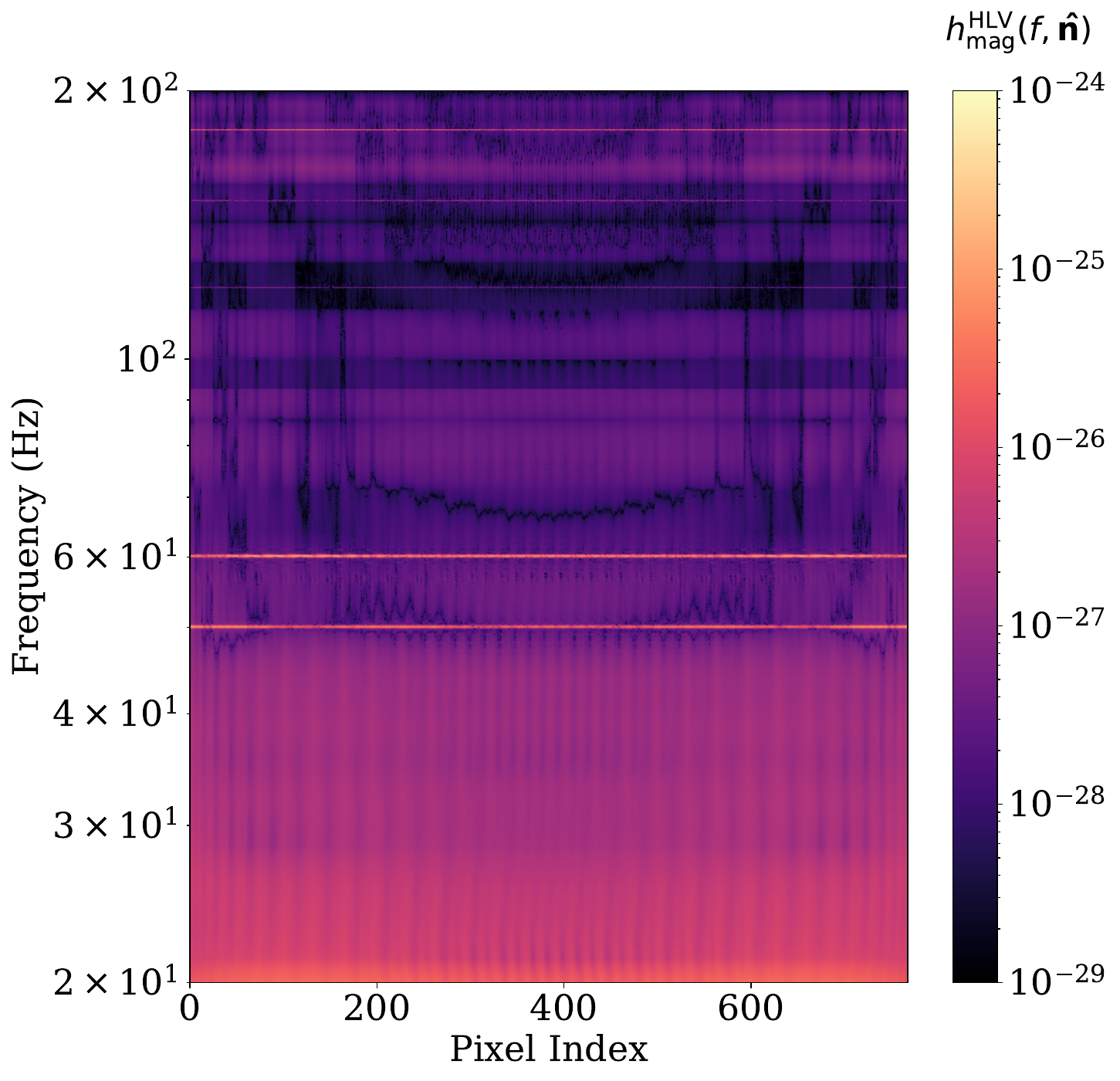} 
        \label{comparison}
    \end{minipage}
    \hfill
    \begin{minipage}[]{0.48\textwidth}
        \centering
        \includegraphics[width=\textwidth,height=3.5in]{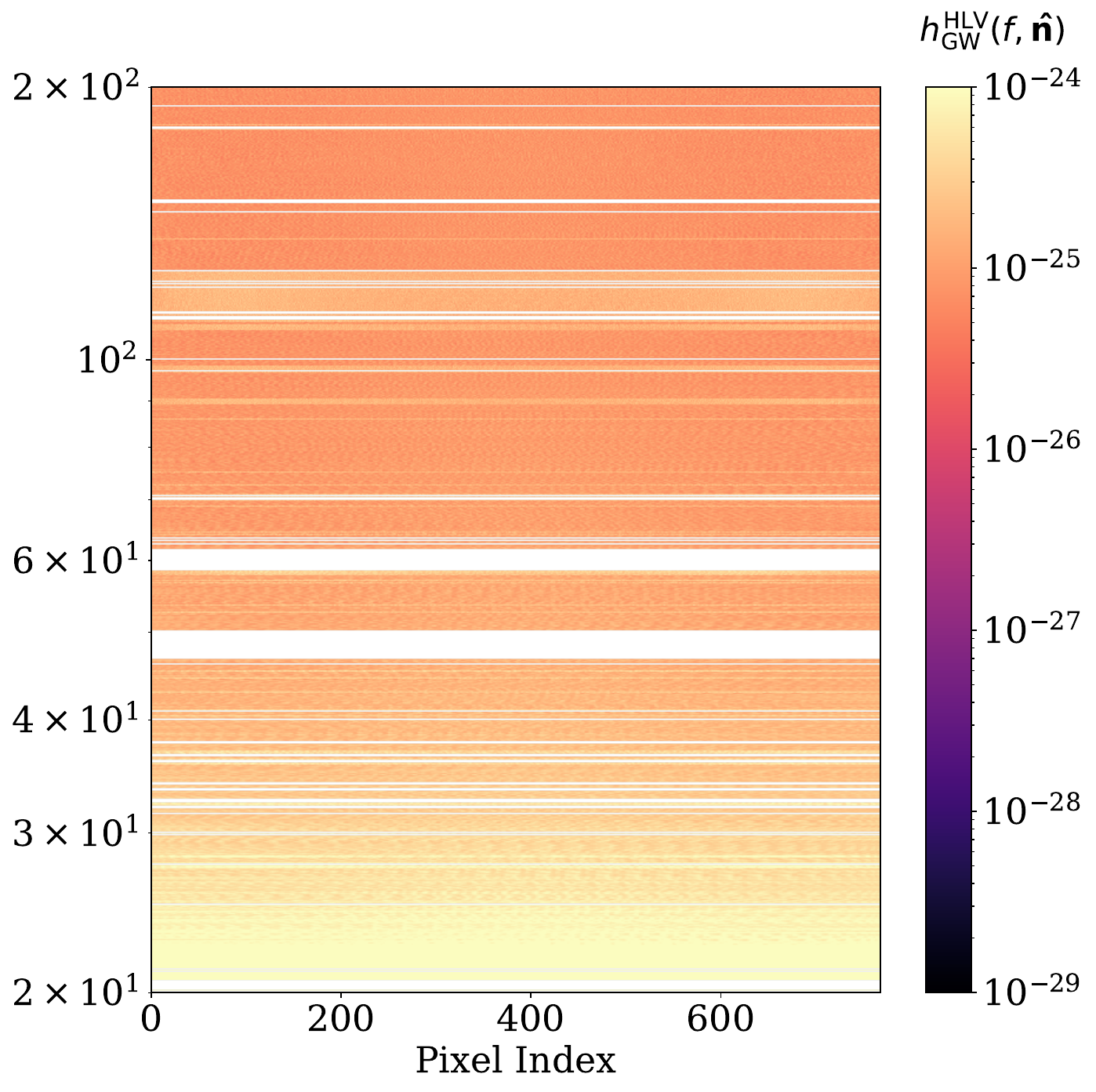} 
        \label{comparison}
    \end{minipage}
   \caption{The effective GW strain $h_{\rm mag}(f,\boldsymbol{\hat{n}})$ related to correlated magnetic noise (left panel) and the 95\% confidence Bayesian upper limits from \cite{KAGRA:2021rmt} on the GW strain $h_{\rm gw}(f,\boldsymbol{\hat{n}})$ amplitude for each sky direction in the 20-200 Hz
    band (right panel). Both plots
    are derived with data from the O3 run, combining the HL, HV, and LV baselines.}
    \label{comparison}
\end{figure*}

\begin{figure}
\includegraphics[width=20pc]{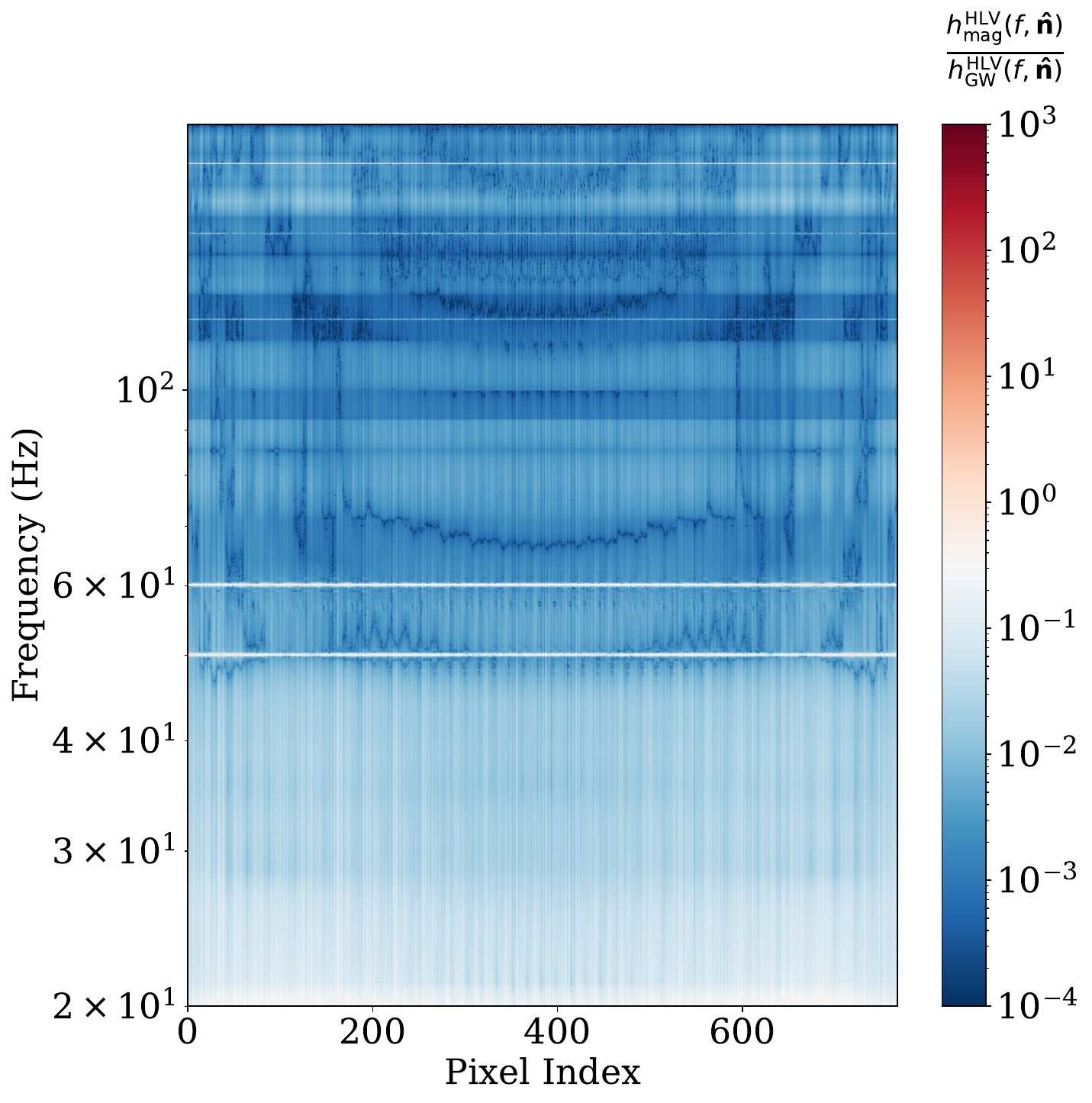}
\caption{The ratio of the effective GW strain, $h_{\rm mag}(f,\boldsymbol{\hat{n}})$, associated with correlated magnetic noise, to the 95\% confidence Bayesian upper limits from \cite{KAGRA:2021rmt} on the GW strain $h_{\rm gw}(f,\boldsymbol{\hat{n}})$ amplitude, for each sky direction in the 20-200 Hz frequency range. } 
\label{ASAF_RATIO_HLV}
\end{figure}

\label{sec:results}
We present the results about the impact of correlated magnetic noise on different SGWB searches during the third observing run \cite{LIGO:2021ppb, KAGRA:2021rmt, KAGRA:2021mth,KAGRA:2021kbb} of the LVK collaboration. We start by illustrating the results for the ASAF search \cite{KAGRA:2021rmt} in subsection \ref{subsec:ASAF_results}, which we then use to derive those for the broad-band radiometer search \cite{KAGRA:2021mth} in subsection \ref{subsec:BBR_results} and the isotropic search \cite{KAGRA:2021kbb} in subsection \ref{subsec:iso_results}. Throughout this section, we pixelate the sky using the Hierarchical Equal Area isoLatitude Pixelization (HEALPix) \cite{Gorski:2004by} scheme with parameter $N_{\rm side}=8$. This results in $N_{\rm pixel}=12 N_{\rm side}^2=768$ pixels, with each pixel having an area approximately equal to $54.7 \rm{deg}^2$. The choice of pixel resolution in this analysis is consistent with the diffraction limit in the 20-200 Hz frequency band, although it is possible to go beyond this, as pointed out by \citet{Floden:2022scq}. We present results in this frequency band using 1/32 Hz as frequency resolution. This
band is expected to be the most sensitive for an SGWB
following a power-law spectrum \cite{Thrane:2013oya} with a spectral index as in the standard LVK searches.

\subsection{Impact on ASAF directional SGWB searches}
\label{subsec:ASAF_results}

The all-sky all-frequency (ASAF) \cite{KAGRA:2021rmt} directional search is an excess power search for narrow-band anisotropic SGWB at each direction in the sky. 
These signals cannot be detected by either broad-band or narrow-band radiometer searches, as broad-band radiometry focuses on sources that emit GWs over a wide frequency band, while narrow-band radiometry historically targets only high-priority sky locations.

In the ASAF directional search, GW radiometry is performed at each frequency bin separately at a low computational cost. This indicates the suitability of the search
for quasi-monochromatic GW signals from sources such as isolated, rotating, non-axisymmetric neutron stars. If excesses corresponding to frequency-location pairs are identified, they can be followed up using more sensitive searches for quasi-monochromatic GW signals that are based on matched filtering \cite{Knee:2023toa} techniques. As a result, the computational cost of searching for these sources in the whole sky has been reduced to a narrow frequency bins and sky locations.

\begin{figure}
\includegraphics[width=20pc]{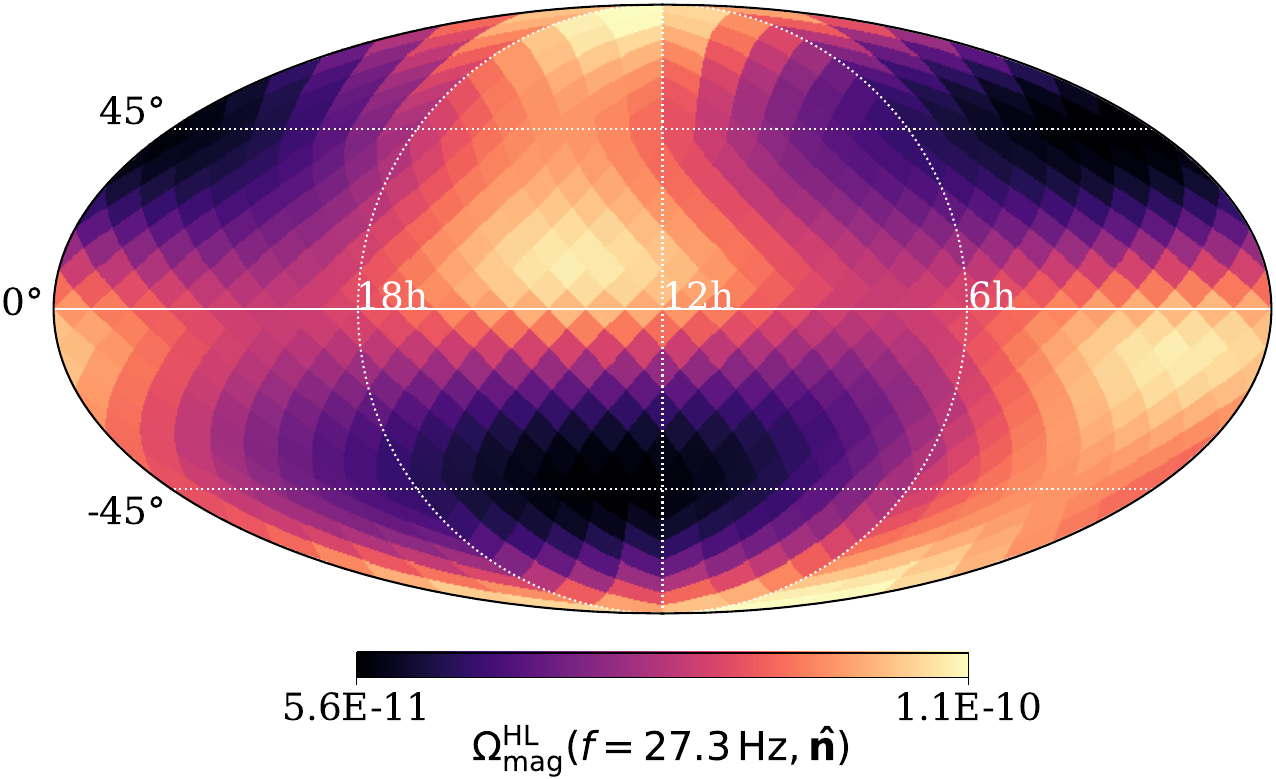}
\caption{The intensity of the effective GW energy density is depicted as a Mollweide projection of the sky in equatorial coordinates at the fourth Schumann resonance peak of 27.3 Hz.} \label{omega_fourth_peak}
\end{figure}

\begin{figure*}
\centering
\includegraphics[width=20pc]{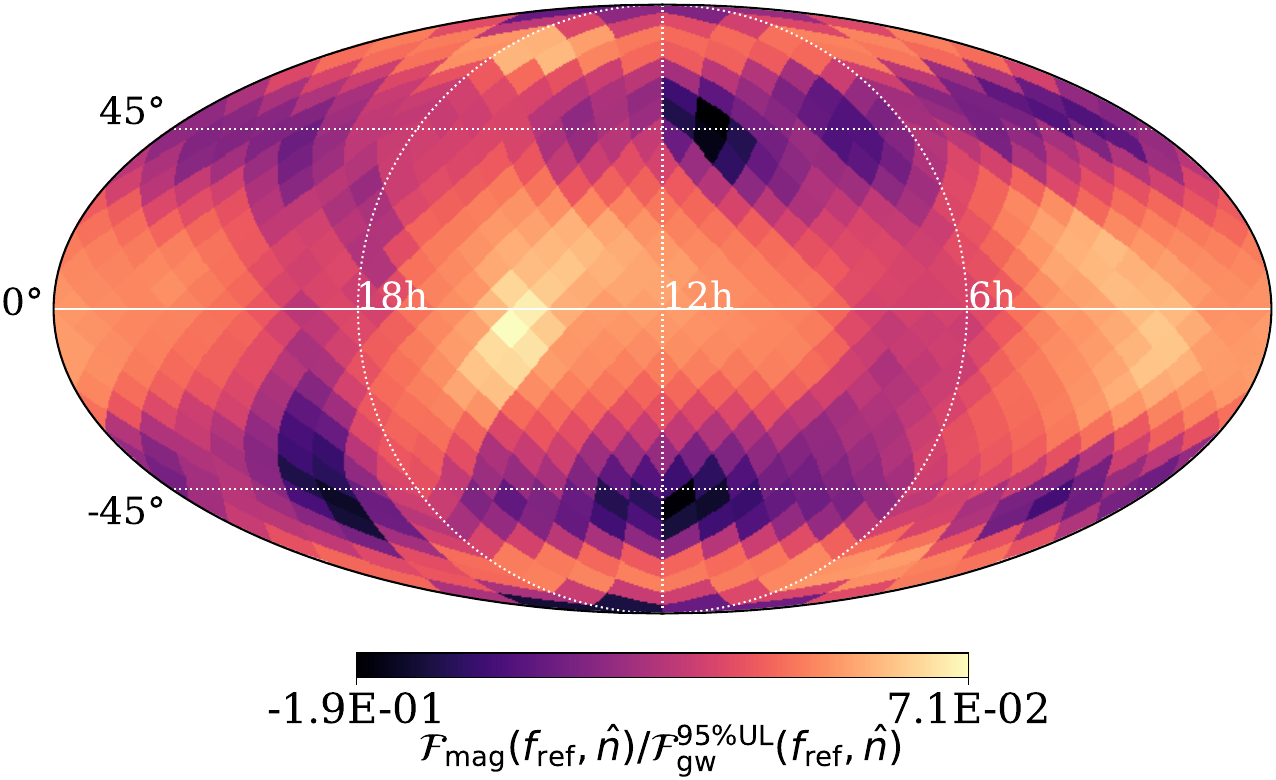}
\caption{Mollweide projection in the equatorial coordinates of the ratio of the effective GW flux of magnetic origin $\mathcal{F}_{\mathrm{mag}}(f_{\rm ref},\boldsymbol{\hat{n}})$ to the BBR upper limits on the GW flux $\mathcal{F}^{95\%, \rm UL}_{\rm gw}(f_{\rm ref},\boldsymbol{\hat{n}})$ using the O3 data sets of the LVK detectors' network.}
\label{broadband}
\end{figure*}

In this sub-section, we examine whether the results from the ASAF O3 run are affected by magnetic correlations. We express the results in terms of an effective GW strain of magnetic origin $h_{\rm mag}$, assuming a circularly polarized signal \cite{KAGRA:2021rmt}.
This strain is defined as
\begin{equation}
\centering
\label{strain}
h_{\rm mag}(f,\boldsymbol{\hat{n}}) = \sqrt{\mathcal{P}_{\rm mag}(f,\boldsymbol{\hat{n}})  \Delta f},
\end{equation}
and it mimics $h_{\rm GW}(f,\boldsymbol{\hat{n}})$ from a monochromatic circularly polarized GW signal \cite{KAGRA:2021rmt}.

In Fig. \ref{comparison}, we present the values of $h_{\rm mag}(f,\boldsymbol{\hat{n}})$ (left panel)
and the $95\%$ Bayesian upper limits on the GW  strain $h^{95\% \rm UL}_{\rm gw}(f,\boldsymbol{\hat{n}})$ (right panel)
from the ASAF O3 search for persistent GWs \cite{KAGRA:2021rmt}. Analogous figures for the HL, HV, and LV baselines are presented in appendix \ref{sec:appendix_ASAF_HL_HV_LV}. From the left panel in Fig. \ref{comparison}, we observe that $h_{\rm mag}(f,\boldsymbol{\hat{n}})$ is most intense in the 20-30 Hz band, where the Schumann resonances are strongest in the LVK band, with $h_{\rm mag}(f,\boldsymbol{\hat{n}})\sim \mathcal{O}(10^{-26}-10^{-25})$ and decreases towards higher frequencies. 
We compare the left and right panels of Fig. \ref{comparison},
by illustrating the ratio $h_{\rm mag}(f,\boldsymbol{\hat{n}})/h^{95\% \rm UL}_{\rm gw}(f,\boldsymbol{\hat{n}})$ in Fig. \ref{ASAF_RATIO_HLV}. We observe that there is no impact of magnetic noise on the O3-HLV ASAF search for persistent GWs, given that $h_{\rm mag}(f,\boldsymbol{\hat{n}})$ is several orders of magnitude below $h^{95\% \rm UL}_{\rm gw}(f,\boldsymbol{\hat{n}})$.

The arch structure in $h_{\rm mag}(f,\boldsymbol{\hat{n}})$ for frequencies above 50Hz is due to the oscillatory nature of the overlap reduction function in Eq. (\ref{orf}). We also observe two intense lines at 50Hz and 60Hz, which are associated with the European and the US power mains, respectively. These two lines and the other frequency bins that are contaminated by noise artifacts (such as mechanical resonances, calibration, and instrumental lines \cite{VIRGO:2012oxz,Virgo:2022ysc,LIGO:2021ppb}) are not included in the O3 ASAF analysis and appear as white bands in $h^{95\% \rm UL}_{\rm gw}(f,\boldsymbol{\hat{n}})$.

We also illustrate the angular distribution of the effective GW energy density for the HL baseline relative to the 27.3Hz peak of the fourth harmonic Schumann resonance spectrum in Fig. \ref{omega_fourth_peak} as a Mollweide projection of the sky. Such an angular power spectrum $\Omega^{\rm HL}_{\rm mag}(f=27.3\,\rm{Hz}$$,\boldsymbol{\hat{n}})$ is of the order $\sim \mathcal{O}(10^{-11}-10^{-10})$. This angular energy spectrum is minor compared with the upper limits of $\Omega^{\rm HL}_{\rm GW}(f=27.3\, \mathrm{Hz}, \boldsymbol{\hat{n}})$, extracted from the ASAF O3 formalism, which are on the order of magnitude of $\sim \mathcal{O}(10^{-6})$. Finally, projections about the impact of magnetic correlations on the ASAF search at the 
A+ design sensitivity are presented in appendix \ref{sec:appendix_A+_projection}.

\subsection{Impact on Broad-band Radiometer searches}
\label{subsec:BBR_results}

Just as the ASAF directional search can be contaminated by this class of noise, specifically due to the Schumann resonance peaks, the BBR directional method \cite{LIGOScientific:2019gaw, KAGRA:2021mth} can also be affected. This is because of the broad-band features of Schumann resonances. 
To estimate the magnitude of this environmental factor in the BBR analysis, we apply a noise-weighted sum to the effective anisotropic estimator $\hat{\mathcal{P}}_{\mathrm{mag}}(f,\, \vu*{n})$ over the entire 20-200 Hz frequency band of interest, namely 
\begin{align}
 \label{ptestbroadband}
 \centering
 &\mathcal{\hat{P}}_{\mathrm{mag}}(f_{\rm ref},\, \boldsymbol{\hat{n}})=\frac{\sum_f \mathcal{\hat{P}}_{\mathrm{mag}}(f,\boldsymbol{\hat{n}}) \sigma_{\boldsymbol{\hat{n}}}^{-2}(f)H_{\rm mag}(f)}{\sum_f \sigma_{\boldsymbol{\hat{n}}}^{-2}(f)H_{\rm mag}^2{(f)}},\\
&\sigma_{\rm mag}(\boldsymbol{\hat{n}})=\left[\sum_f \sigma_{\boldsymbol{\hat{n}}}^{-2}(f)H_{\rm mag}^2{(f)} \right]^{-1/2},
\end{align}
where 
\begin{equation}
\centering
\label{spectral_shape}
H_{\rm mag}(f)=\frac{\Omega_{\rm mag}(f)}{\Omega_{\rm mag}(f_{\rm ref})}\left(\frac{f}{f_{\rm ref}}\right)^{-3},
\end{equation}
with $f_{\rm ref}=25 {\rm Hz}$.
We then convert $\mathcal{P}_{\rm mag}(f_{\rm ref},\boldsymbol{\hat{n}})$ to
an effective magnetic flux $\mathcal{F}(f_{\rm ref},\boldsymbol{\hat{n}})$, by using the following relation  \cite{KAGRA:2021mth}

\begin{equation}
\centering
\mathcal{F}_{\mathrm{mag}}(f_{\rm ref},\, \boldsymbol{\hat{n}})=\frac{c^3 \pi}{4G}f_{\rm ref}^2 \mathcal{P}_{\mathrm{mag}}(f_{\rm ref},\boldsymbol{\hat{n}}).
\end{equation}

In Fig. \ref{broadband}, we present the ratio of $\mathcal{F}_{\rm mag}(f_{\rm ref},\boldsymbol{\hat{n}})$ based on O3 magnetic data from the HLV network
and the $95\%$ Bayesian upper limits on the GW flux $\mathcal{F}^{95\%, \rm UL}_{\rm gw}(f_{\rm ref},\boldsymbol{\hat{n}})$ from the LVK O3 BBR search \cite{KAGRA:2021mth}.
The denominator (GW flux) corresponds to a flat power spectral density (labelled with $\alpha=3$ in \cite{KAGRA:2021mth}). The map uses a Mollweide projection in equatorial coordinates, with the flux expressed in units of $\mathrm{erg\,  cm^{−2}\, s^{-1}\, Hz^{−1} }$.

Comparing these results, we observe that the magnetic correlations do not significantly affect the O3 BBR searches, given that there is at least one order of magnitude difference between them and the O3 LVK upper limits. Specifically, the effective quantity appears as $\mathcal{F}_{\mathrm{mag}}(f_{\rm ref},\boldsymbol{\hat{n}})\sim\mathcal{O}{(10^{-12}-10^{-11})}$ at each sky direction, in comparison with $\mathcal{F}^{95\%, \rm UL}_{\rm gw}(f_{\rm ref},\boldsymbol{\hat{n}})\sim$$\mathcal{O}{(10^{-10}-10^{-9})}$.

\subsection{Impact on Isotropic SGWB searches}
\label{subsec:iso_results}

Searches for the isotropic SGWB \cite{KAGRA:2021kbb} can also be affected by magnetic correlated noise.
In this section, we evaluate the magnetic budget for these analyses, utilizing the results extracted in Section \ref{subsec:ASAF_results}, and compare them with the ones from \citet{Janssens:2022tdj}.

\begin{figure*}[]
\centering
\includegraphics[width=40pc]{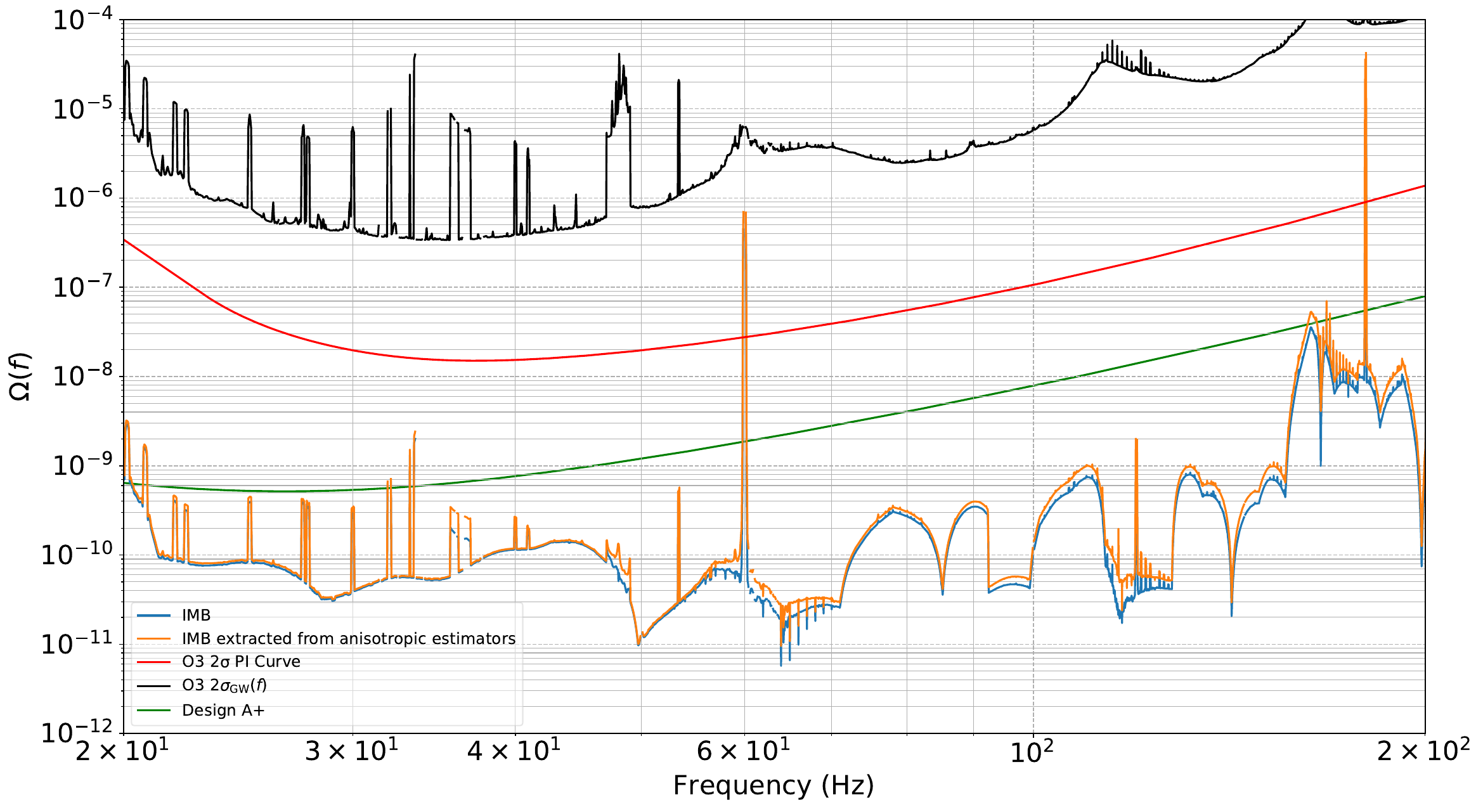}
\caption{Comparison of the IMB from \citet{Janssens:2022tdj} expressed with the blue line and the IMB derived from the effective GW anisotropic estimators (orange line). In both analyses, the IMBs have been constructed for the HLV detector network. The black and red curves indicate the narrow-band and broad-band sensitivities for the O3 run, respectively \cite{KAGRA:2021kbb}. Lastly, the green curve represents the Design A+ broad-band sensitivity \cite{A+}.}
\label{omega_total_iso}
\end{figure*}

The estimator of the isotropic strain spectral density $\mathcal{\hat{P}}_{\rm iso, gw}(f)$ can be obtained as
\begin{equation}
\centering
\label{iso_estimator}
\mathcal{\hat{P}}_{\rm iso}\sigma_{\rm iso}^{-2}(f)=\frac{5}{4\pi}\int d\boldsymbol{\hat{n}}\mathcal{\hat{P}}_{\mathrm{gw}}(f,\boldsymbol{\hat{n}})\sigma_{\boldsymbol{\hat{n}}}^{-2}(f),
\end{equation}
with the relative uncertainty 
\begin{equation}
\label{iso_estimator_uncert}
\sigma_{\rm iso}^{-2}(f)=\left(\frac{5}{4\pi}\right)^{2}\int d\boldsymbol{\hat{n}} \int d \boldsymbol{\hat{n}'}\Gamma_{\boldsymbol{\hat{n}}\boldsymbol{\hat{n}}'}(f),
\end{equation}
including the integration over the sky directions of the Fisher matrix. In both equations, given that the HEALPix provides a discrete grid, the integration over all the sky directions has been performed using $d\boldsymbol{\hat{n}}=4\pi/ N_{\rm pix}$.
In an equivalent way, we evaluate the $\mathcal{\hat{P}}_{\rm iso, mag}(f)$, replacing $\mathcal{\hat{P}}_{\mathrm{gw}}(f,\boldsymbol{\hat{n}})$ with $\mathcal{\hat{P}}_{\mathrm{mag}}(f,\boldsymbol{\hat{n}})$ in Eq. (\ref{iso_estimator}). 
To compare the results with the ones in \citet{Janssens:2022tdj}, we recast $\mathcal{\hat{P}}_{\rm iso, mag}(f)$ to $\Omega_{\rm mag}(f)$ using the Eq. \eqref{Omega_mag}
and plot the effective spectrum in the 20-200 Hz band in Fig. \ref{omega_total_iso}.

Fig. \ref{omega_total_iso} illustrates the isotropic magnetic budget (IMB) for the HLV detector network, indicated by the blue line from \citet{Janssens:2022tdj}, as well as the IMB extracted from anisotropic estimators (orange line). The two lines determine the upper range of the IMB. The black curve indicates the square root of the variance \cite{KAGRA:2021kbb} for isotropic SGWB searches based on the O3 LVK observing run, describing the sensitivity of the method to narrow-band features. The red curve
describes the broad-band sensitivity of the search to SGWBs with power-law form, known as the PI curve \cite{Thrane:2013oya}. Lastly, the green line denotes the Design A+ broad-band sensitivity.

Observing Fig. \ref{omega_total_iso}, the extracted magnetic estimators starting from the directional analysis agree to a good extent with those from the isotropic analysis. 
The differences between the two plots appear due to the discrete integration of the overlap reduction function over the sky in Eq. (\ref{iso_estimator_uncert}) and the choice of constant parameter $N_{\rm side}$ over the whole frequency range. The narrow-band peaks of the Schumann resonances in the 20-60 Hz range are well recovered in both analyses. Other significant fluctuations (peaks and dips) in the spectrum arise due to the high-frequency behaviour of the (sparsely sampled) ITC coupling function for the LIGO detectors (sitewide coupling) \cite{AdvLIGO:2021oxw}.

Given the compatibility of the results with \citet{Janssens:2022tdj}, we confirm once more that Schumann resonances did not appreciably affect the searches for isotropic SGWB during the O3 observing run. We extract this conclusion from the appearance of both lines below the narrow-band sensitivity (black line). We remark that we did not include any uncertainties on the magnetic coupling functions. As it was pointed out in \citet{Janssens:2022tdj}, there is an overlap between the 2-3$\sigma$ IMB and the A+ design sensitivity curve, indicating potential magnetic noise contamination once the detectors reach their
design configurations (A+ curve in the Fig.  \ref{omega_total_iso}).

\section{Conclusions}
Correlated magnetic noise can couple to the GW detectors, specifically within the frequency band of 20-200 Hz,
inducing an effective GW strain. This effective GW strain could potentially bias the detection statistics of the SGWB, appearing in the cross-correlation of the detector output. This study investigates for the first time the impact of this environmental factor on anisotropic SGWB searches, using data from magnetometers at each detector site during the O3 observing run. In light of its exploratory nature, we have neglected the phase of the couplings between the magnetic fields and the GW interferometers, as its interplay with the overlap reduction function in SGWB searches is non-trivial. This will be subject of future works.

\begin{figure*}
\centering
\begin{minipage}{.5\linewidth}
  \centering
  \includegraphics[width=\textwidth,height=3in]{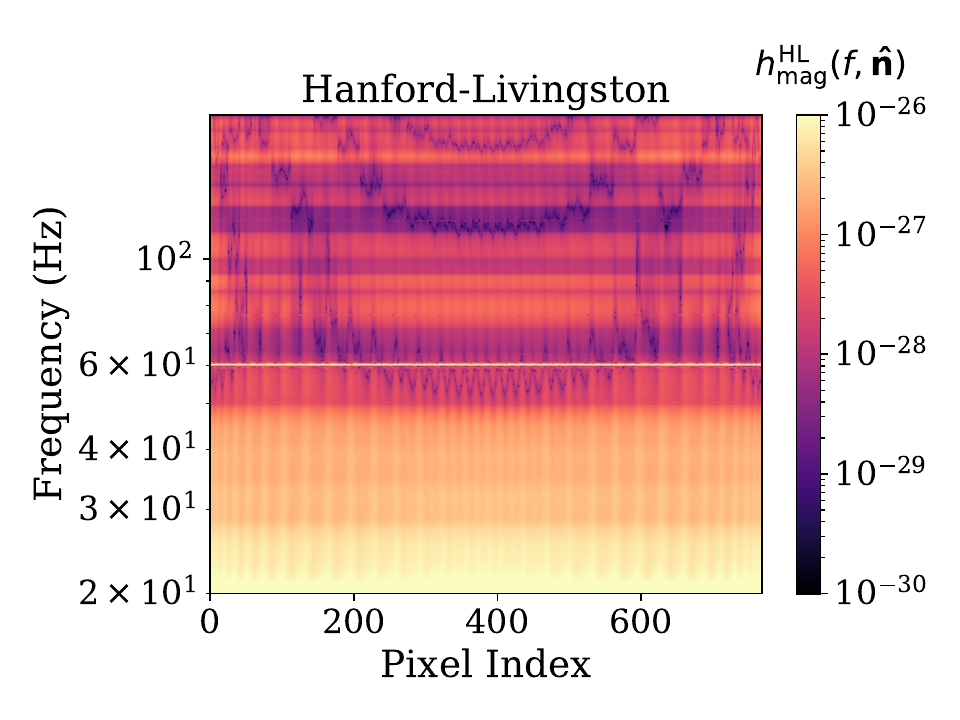}
\end{minipage}%
\begin{minipage}{.5\linewidth}
  \centering
  \includegraphics[width=\textwidth,height=3.in]{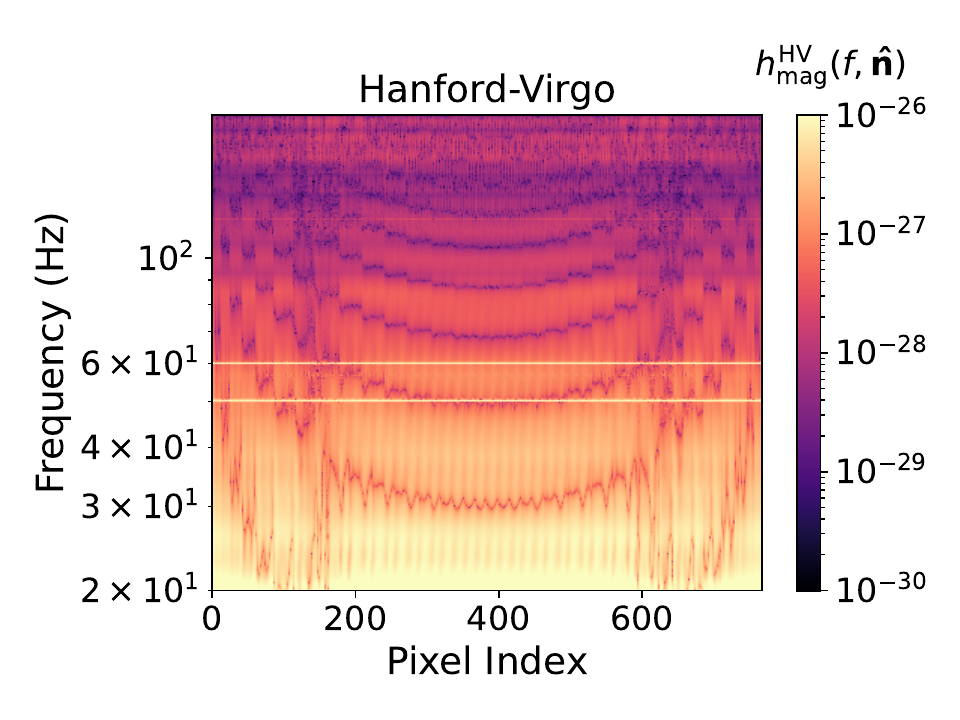}
\end{minipage}
\begin{minipage}{.5\linewidth}
  \centering
  \includegraphics[width=\textwidth,height=3.in]{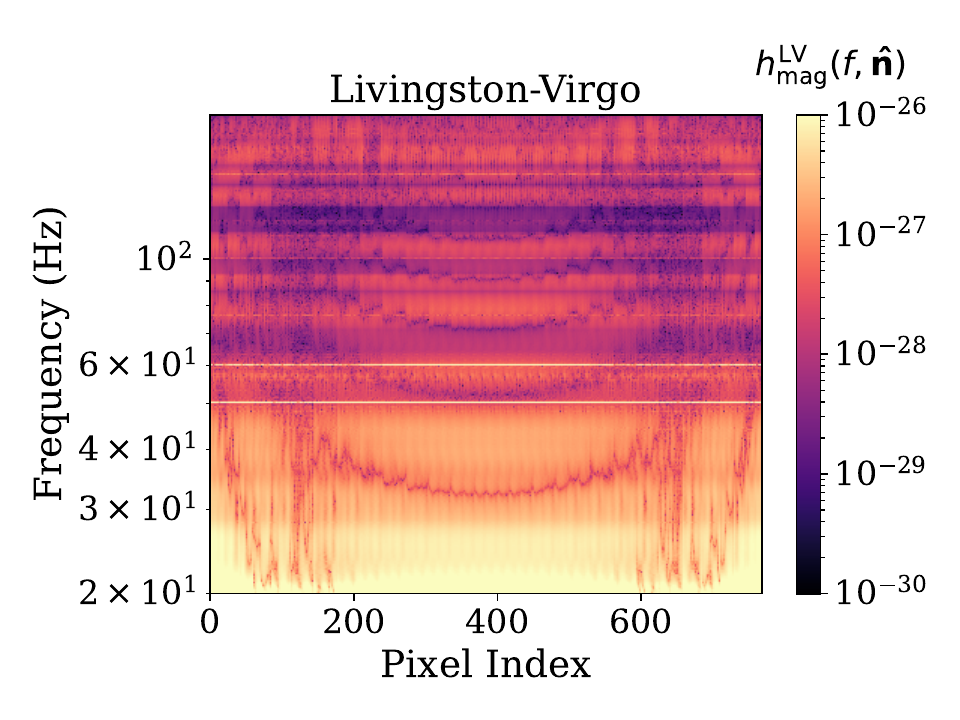}
\end{minipage}
\captionof{figure}{Effective GW strain $h_{\mathrm{mag}}(f,\boldsymbol{\hat{n}})$ for HL, HV, and LV baselines of the LVK detector network.}
\label{ASAF_individual}
\end{figure*}

We estimated the impact of correlated magnetic noise on the ASAF directional search for the LVK detector network \cite{KAGRA:2021rmt}.
It was found that the effective strain $h_{\rm mag}(f,\boldsymbol{\hat{n}})$ manifests 
intensely in the 20-30 Hz frequency band in the orders of magnitude of $h_{\rm mag}(f,\boldsymbol{\hat{n}})\sim \mathcal{O}(10^{-26}-10^{-25})$. This effective strain is small in comparison with the $95\%$ Bayesian strain upper limit $h^{95\% \rm UL}_{\rm gw}(f,\boldsymbol{\hat{n}})$ from O3 data. Thus, the comparison indicates that in each sky direction and frequency bin, the magnetic correlations did not affect the ASAF directional search of the O3 run. The results for the HL, HV, and LV individual baselines are presented in appendix \ref{sec:appendix_ASAF_HL_HV_LV}.

We also investigated the scenario of possible magnetic contamination of the BBR results from O3 analysis. After converting the BBR effective estimator to the effective flux $\mathcal{F}_{\rm mag}(f_{\rm ref},\boldsymbol{\hat{n}})$, we compared this quantity with the $95\%$ Bayesian upper limits on the GW flux $\mathcal{F}^{95\%, \rm UL}_{\rm gw}(f_{\rm ref},\boldsymbol{\hat{n}})$ for all the baselines combined. We conclude that the O3 BBR search was unaffected by this environmental correlated factor.

We also extracted the isotropic effective estimator $\mathcal{\hat{P}}_{\rm iso, mag}(f)$ after integrating discretely over all the directions in the sky. The produced magnetic budget agrees to a good extent with that in \citet{Janssens:2022tdj}, indicating the validation of the analysis. Both IMBs lie below the narrow-band sensitivity curve of O3 SGWB searches, confirming the absence of magnetic contamination.

We conclude that, up to the O3 observing run, both searches for isotropic and anisotropic SGWBs were unaffected by correlated magnetic noise. 
However, the possibility that Schumann resonances could limit the sensitivity of terrestrial detectors to SGWB in future observing runs remains. 
In this context, we explored the scenario of contamination of directional searches for SGWB from correlated magnetic noise at the LIGO A+ design sensitivity levels in appendix \ref{sec:appendix_A+_projection}, after extracting the PI-like sensitivity curve \ref{sec:appendix_PI_curve} for the ASAF radiometer search. We found that there is no magnetic contamination, apart from a few frequency bins around 60 Hz and 180 Hz.

\begin{acknowledgments}
This material is based upon work supported by NSF's LIGO Laboratory which is a major facility fully funded by the National Science Foundation. The LIGO document number of this article is LIGO-P2400508.

The authors would like to express their gratitude to Vuk Mandic and Nick van Remortel for their detailed reviews of the manuscript and valuable comments.

S.V. and F.D.L. are supported by a 
Grant, Fonds pour la formation à la Recherche dans l'Industrie et dans l'Agriculture (FRIA) from Belgian Fundings for Research, provided by F.R.S.-FNRS (Fonds de la Recherche Scientifique - FNRS).
K.J. was supported by FWO-Vlaanderen via grant number 11C5720N during part of this work.

In this work, we used \texttt{numpy} \cite{vanderWalt:2011bqk} and \texttt{scipy} \cite{Virtanen:2019joe} packages for the data analysis. We also used \texttt{matplotlib} \cite{Hunter:2007ouj} package to produce the Figs. \ref{comparison}, \ref{omega_total_iso}, \ref{ASAF_individual}, \ref{A_PLUS_FIGURES} and \texttt{healpy} \cite{Zonca2019} for the maps in Figs. \ref{omega_fourth_peak}, \ref{broadband}.

\end{acknowledgments}

\appendix

\section{Individual impact on ASAF directional search for HL, HV, and LV detector baselines}
\label{sec:appendix_ASAF_HL_HV_LV}

We present the effective GW strain $h_{\mathrm{mag}}(f,\boldsymbol{\hat{n}})$ for each detector baseline of the LVK network in Fig. \ref{ASAF_individual}. The Schumann resonance peaks produce intense effective magnetic strain
$h_{\rm mag}(f,\boldsymbol{\hat{n}})\sim \mathcal{O}(10^{-25})$ at each individual baseline HL, HV and LV in the 20-25Hz band and $h_{\rm mag}(f,\boldsymbol{\hat{n}})\sim \mathcal{O}(10^{-26})$ in the 25-30Hz frequency range. Towards higher frequencies, in the range of
30-60 Hz, the induced GW strain is $h_{\rm mag}(f,\boldsymbol{\hat{n}})\sim \mathcal{O}(10^{-27}-10^{-26})$
for all the baselines.
Lastly, for frequencies above 60 Hz, the possibility of magnetic contamination decreases as the impact of correlated magnetic fields appears at the order of magnitude of $h_{\rm mag}(f,\boldsymbol{\hat{n}})\sim \mathcal{O}(10^{-28}-10^{-27})$.

\section{Impact of magnetic correlations on LIGO detectors at A+ design sensitivity}
\label{sec:appendix_A+_projection}

\begin{figure*}
\centering
\begin{minipage}{.5\linewidth}  
  \centering
\includegraphics[width=\textwidth,height=3.15in]{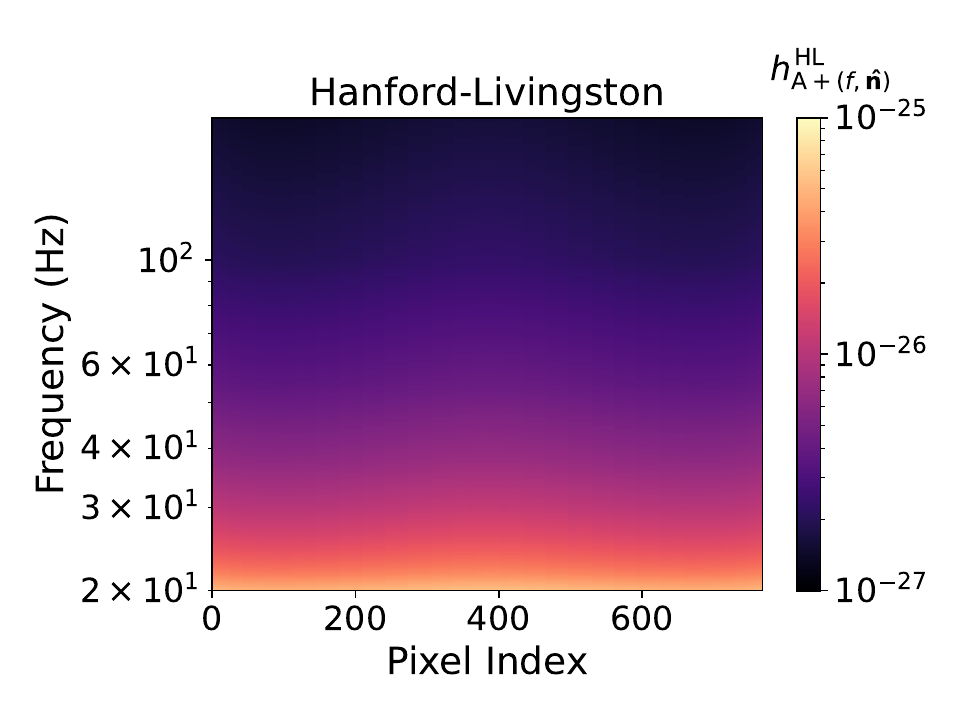}
\end{minipage}%
\begin{minipage}{.55\linewidth}  
  \centering
\includegraphics[width=\textwidth,height=3.15in]{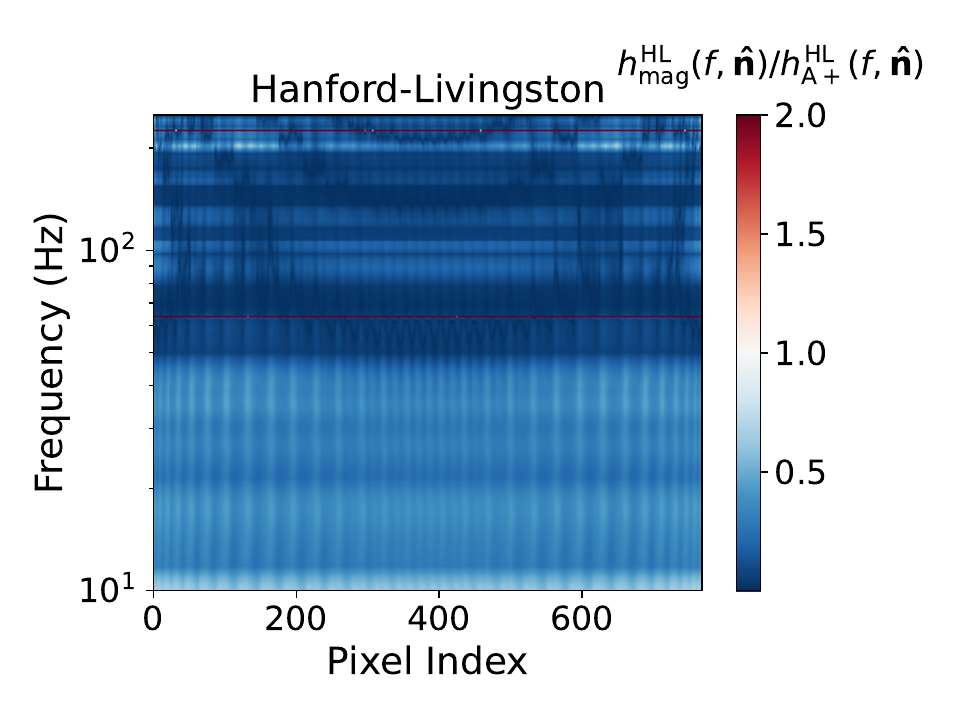}
\end{minipage}
\captionof{figure}{Projected GW strain from LIGO detectors at A+ design sensitivity (left plot) and ratio of magnetic correlations measured during the O3 run to the projected GW strain for the HL detector baseline (right plot).}
\label{A_PLUS_FIGURES}
\end{figure*}

In the near future, the LIGO terrestrial interferometers are expected to reach the A+ design sensitivities \cite{A+}. 
With the sensitivity improvement, the presence of correlated magnetic noise in the detectors' data streams could no longer be neglected when performing directional searches for anisotropic SGWBs. To explore such a scenario, we present the projections for the sensitivity of the ASAF radiometer search and compare them with the corresponding magnetic budget we presented in appendix \ref{sec:appendix_ASAF_HL_HV_LV}. 

The derivation of the sensitivity curve for radiometer searches follows the methods described in appendix \ref{sec:appendix_PI_curve}. We illustrate such a projection in the left panel of Fig. \ref{A_PLUS_FIGURES}, where we consider the Hanford-Livingston baseline A+ power spectral densities for a one-year observing time and an SNR equal to 2. For visualization ease, we illustrate in the right panel of Fig. \ref{A_PLUS_FIGURES} the ratio between the O3 ASAF effective GW strain of magnetic origin from the top left panel of Fig. \ref{ASAF_individual} in the appendix \ref{sec:appendix_ASAF_HL_HV_LV} and the projected sensitivity for $h_{\rm gw}(f)$ of the A+ HL baseline. If this ratio is larger than one, it means that correlated magnetic noise is likely no longer negligible when performing directional SGWB searches. We conclude that the ratio is less than 1 for all the frequencies bin, with the exception of the 60-Hz and 180-Hz ones, which correspond to the two narrow-band peaks of the isotropic magnetic budget in Fig. \ref{omega_total_iso}.

\section{Derivation of the PI-like sensitivity curves for SGWB radiometer searches in the absence of pixel-to-pixel correlations}
\label{sec:appendix_PI_curve}
In this appendix, we sketch the derivation of the formula we have employed to produce the projections of the sensitivity of A+ LIGO detectors \cite{A+} for SGWB radiometer searches. Given the absence of such sensitivity curves in the literature, we derive them here for the first time, taking inspiration from the power-law integrated (PI) sensitivity curves that exist for searches for isotropic SGWB \cite{Thrane:2013oya}. Throughout the derivation, we make three assumptions, namely that 1) we are performing a radiometer analysis, 2) we neglect pixel-to-pixel correlations, and 3) we can approximate detectors' noise power spectral densities to be constant over time.  

\begin{widetext}
    The radiometer analysis method assumes as the signal model a point source in the direction $\vu*{n}_{0}$ and angular power spectral density $\mathcal{P}_{\vu*{n}_{0}}(f)$
    \begin{equation}
        \label{eq:P_f_n_radiometer_for_PI}
        \mathcal{P}(f, \, \vu*{n}) = \mathcal{P}_{\vu*{n}_{0}}(f)\, \delta^{2}(\vu*{n} - \vu*{n}_{0}).
    \end{equation}
    As a consequence, the expectation value of the cross-correlation statistic ${\rm CSD }(t;\,f) \equiv \tilde{s}^{*}_{I}(t;f)\tilde{s}_{J}(t;f)$ becomes
    \begin{equation}
        \label{eq:CC_radiometry_expectation_value_for_PI}
        \expval{{\rm CSD}_{IJ} (t; \, f)} = \frac{\tau}{2}\int_{\mathcal{S}_{2}} \dd[2]{\vu*{n}} \gamma_{IJ}(t; \, f, \, \vu*{n})\,  \mathcal{P_{\rm gw}}(f, \, \vu*{n}) = \frac{\tau}{2} \gamma_{IJ}(t; \, f,\, \vu*{n}_{0})\,  \mathcal{P}_{\vu*{n}_{0}}(f),
    \end{equation}
     where one has used Eq. \eqref{eq:P_f_n_radiometer_for_PI} in the second equality, with $\tau$ and the geometrical factor $\gamma_{IJ}(t; \, f\, \vu*{n})$ having the same meaning as in the main text.

     The signal-to-noise ratio ${\rm SNR}_{IJ}(f,\, \vu*{n})$ of a baseline $IJ$ for the radiometer search in the pixel basis is given by
     \begin{equation}
         \label{eq:SNR_radiometer_for_PI}
         {\rm SNR}(f,\, \vu*{n}) = \frac{\hat{\mathcal{P}}_{\rm gw} (f, \, \vu*{n})}{\sigma_{\hat{\mathcal{P}}_{\rm gw}}(f,\, \vu*{n})} = \frac{[\Gamma^{-1}_{IJ}]_{\vu*{n} \vu*{n'}}(f) \vdot [X_{IJ}]_{\vu*{n'}}(f)}{\sqrt{{\rm diag}[\Gamma^{-1}_{IJ}]_{\vu*{n} \vu*{n}}(f)}} \approx \frac{X_{IJ}(f,\, \vu*{n})}{\sqrt{\Gamma_{IJ}(f,\, \vu*{n})}},
     \end{equation}
where, in the last equality, we have made use of the absence of pixel-to-pixel correlation and hence approximated the Fisher matrix to be diagonal. By making use of the dirty map and (diagonal of the) Fisher matrix definitions in Eqs. \eqref{Dirty}, \eqref{Fisher} in the main text and assuming the detectors' noise to be constant over time, one can recast the above equation as
\begin{equation}
    \label{eq:SNR_radiometer_constant_PSDs}
    {\rm SNR}(f,\, \vu*{n}) = 2\sqrt{\frac{2\Delta f/\tau}{P_{{I}}(f)P_{{J}}(f)}}\, \frac{\sum_{t} \gamma_{IJ}^{*}(t; \, f, \, \vu*{n})\, {\rm CSD }(t;\,f)}{\sqrt{\sum_{t} \gamma_{IJ}^{*}(t; \, f, \, \vu*{n})\, \gamma_{IJ}(t; \, f, \, \vu*{n})}}.
\end{equation}
Then, by considering the expectation value of the above equation and making use of Eq. \eqref{eq:CC_radiometry_expectation_value_for_PI}, one gets
\begin{equation}
    \label{eq:SNR_narrowband_PI}
    {\rm SNR}_{\vu*{n}_{0}}(f) = \sqrt{2\tau\Delta f}\, \sqrt{\frac{\sum_{t} \gamma_{IJ}^{*}(t; \, f, \, \vu*{n})\, \gamma_{IJ}(t; \, f, \, \vu*{n})\, \mathcal{P}^{2}_{\vu*{n}_{0}}(f)}{P_{{I}}(f)P_{{J}}(f)}},
\end{equation}
which can be inverted and eventually give a narrow-band, all-sky, all-frequency sensitivity curve for a given SNR and an observation time $T_{\rm obs}$ divided in chunks of duration $\tau$ labelled by $t$
\begin{equation}
    \label{eq:P_gw_ASAF_PI}
    \mathcal{P}_{\vu*{n}_{0}}(f) = \frac{{\rm SNR}_{\vu*{n}_{0}}(f)}{\sqrt{2\tau\Delta f}}\, \left[\frac{\sum_{t} \gamma_{IJ}^{*}(t; \, f, \, \vu*{n})\, \sum_{t} \gamma_{IJ}(t; \, f, \, \vu*{n})}{P_{{I}}(f)P_{{J}}(f)}\right]^{-1/2}.
\end{equation}

Additionally, if the model of interest for the analysis assumes a broad-band signal, one can take advantage of it and combine the information from different frequency bins. In practice, one can sum in quadrature the narrow-band SNRs from Eq. \eqref{eq:SNR_narrowband_PI}, which becomes an integral in the limit $\Delta f \rightarrow \dd{f}$ and yields the broad-band SNR
\begin{equation}
    \label{eq:SNR_broadband_PI}
    {\rm SNR}_{\vu*{n}_{0}} = \sqrt{2\tau}\, \sqrt{\int_{f_{\rm min}}^{f_{\rm max}} \dd{f} \frac{\mathcal{P}^{2}_{\vu*{n}_{0}}(f)\, \sum_{t} \gamma_{IJ}^{*}(t; \, f, \, \vu*{n})\, \gamma_{IJ}(t; \, f, \, \vu*{n})}{P_{{I}}(f)P_{{J}}(f)}}.
\end{equation}
To build a PI-like sensitivity curve for $\mathcal{P}_{\vu*{n}_{0}}(f)$, one must further assume that the spectral shape of the signal follows a power-law in frequency
\begin{equation}
    \label{eq:PL_spectrum_for_PI}
    \mathcal{P}_{\vu*{n}_{0}}(f) = P_{\rm ref, \, \alpha,\, \vu*{n}_{0}}\, \left(\frac{f}{f_{\rm ref}}\right)^{\alpha - 3} \equiv P_{\rm ref, \, \alpha,\, \vu*{n}_{0}}\, H_{\alpha}(f),
\end{equation}
plug it in Eq. \eqref{eq:SNR_broadband_PI} and invert it in favour of $\mathcal{P}_{\rm ref, \, \alpha,\, \vu*{n}_{0}}$, eventually obtaining
\begin{equation}
    \label{eq:P_gw_PI_curve}
    \mathcal{P}_{\rm ref, \, \alpha,\, \vu*{n}_{0}} = \frac{{\rm SNR}_{\vu*{n}_{0}}}{\sqrt{2\tau}}\, \left[\int_{f_{\rm min}}^{f_{\rm max}} \dd{f} \frac{H_{\alpha}^{2}(f)\, \sum_{t} \gamma_{IJ}^{*}(t; \, f, \, \vu*{n})\, \gamma_{IJ}(t; \, f, \, \vu*{n})}{P_{{I}}(f)P_{{J}}(f)}\right]^{-1/2}.
\end{equation}
One can observe the similarities between this sensitivity curve and the one for the searches for isotropic SGWBs that was proposed in \cite{Thrane:2013oya}, where $\mathcal{P}_{\rm ref, \, \alpha,\, \vu*{n}_{0}}$ is replaced by $\Omega_{\beta}$, $H_{\alpha}(f)$ by $(f/f_{\rm ref})^{\beta}$, and $\sum_{t} \gamma_{IJ}^{*}(t; \, f, \, \vu*{n})\, \gamma_{IJ}(t; \, f, \, \vu*{n})$ by $\Gamma_{IJ}^{2}(f)$.
To obtain the PI-like directional curve for $\mathcal{P}_{\vu*{n}_{0}}(f)$, one can follow the same steps as in \cite{Thrane:2013oya} and define
\begin{equation}
    \label{eq:actual_P_gw_PI_curve}
    \mathcal{P}_{\vu*{n}_{0}\, PI}(f) \equiv \max_{\alpha} \left[\mathcal{P}_{\rm ref, \, \alpha,\, \vu*{n}_{0}} H_{\alpha}(f)\right].
\end{equation}
We use this all-sky, all-frequency PI-like curve (which we convert to $h_{\rm gw}(f)$ units using Eq.\eqref{strain}) to make predictions about the sensitivity of the detector network in this work.

\end{widetext}

\bibliographystyle{unsrtnat}

\bibliography{citations}

\end{document}